\documentstyle[11pt,aaspp4]{article}
\begin{document}
%\draft
\title{Angular size and emission time scales of relativistic fireballs}
\author{Eli Waxman}
\affil{Institute for Advanced Study, Princeton, NJ 08540; E-mail:
waxman@sns.ias.edu}
%\date{\today}

\begin{abstract}

The detection of delayed X-ray, optical and radio emission, ``afterglow,'' 
associated with $\gamma$-ray bursts (GRBs) is consistent with models, where the
bursts are produced by relativistic expanding blast waves, driven by expanding 
fireballs at cosmological distances. In particular, the time scales over which 
radiation is observed at different wave bands agree with model predictions. It
had recently been claimed that the commonly used relation between observation 
time $t$ and blast wave radius $r$, $t=r/2\gamma^2(r)c$ where $\gamma$ is the 
fluid Lorentz factor, should be replaced with $t=r/16\gamma^2(r)c$ due to blast
wave deceleration. Applying the suggested deceleration modification would make 
it difficult to reconcile observed time scales with model predictions. It would
also imply an apparent source size which is too large to allow attributing 
observed radio variability to diffractive scintillation. We present a detailed 
analysis of the implications of the relativistic hydrodynamics of expanding 
blast waves to the observed afterglow. We find that modifications due to shock
deceleration are small, therefore allowing for both the observed afterglow time
scales and for diffractive scintillation. We show that at time $t$ the fireball
appears on the sky as a narrow ring of radius $h=r/\gamma(r)$ and width 
$\Delta h/h\sim0.1$, where $r$ and $t$ are related by $t=r/2\gamma^2(r)c$.

\end{abstract}
\keywords{gamma rays: bursts}

\section{Introduction}

The availability of accurate positions for GRBs from the BeppoSAX satellite
(\cite{bs1,bs2a,bs2b,bs3}) 
allowed for the first time to detect delayed emission
associated with GRBs in X-ray (\cite{bs1},b,\cite{bs2a,bepx1},b), 
optical (\cite{opt1,opt1a,Paradijs,O6654,O6655}) 
and radio (\cite{R}) wave-bands. The detection of absorption lines in the
optical afterglow of GRB970508 provided the first direct estimate of
source distance, constraining the redshift of GRB970508 to $0.8<z<2.3$ 
(\cite{z}). Observed X-ray to radio afterglows are broadly consistent 
with models based on relativistic blast waves at cosmological distances
(\cite{PRAG,MRAG,VAG,AGWa},b,\cite{AGWMR}). Using these models,
combined radio and optical data allowed for the first time 
to directly estimate the total
GRB energy, implying an energy of $\sim10^{52}{\rm erg}$ for GRB970508
(\cite{AGWb}).

In fireball models of GRB afterglow, the energy released by an explosion,
$\sim10^{52}$erg, is converted to kinetic energy of a thin baryon shell 
expanding at ultra-relativistic speed. After producing the GRB, the shell 
impacts on surrounding gas, driving an ultra-relativistic shock 
into the ambient medium. After a short transition phase, the expanding
blast wave approaches a self-similar behavior where the expansion Lorentz 
factor
decreases with radius as $\gamma\propto r^{-3/2}$. The expanding
shock continuously heats fresh gas and accelerates relativistic electrons,
which produce the observed radiation through synchrotron emission.

Photons emitted at radius $r$ with frequency $\nu$, measured in 
the shell rest frame, are observed over a wide frequency range, 
$2\gamma(r)\nu$ to $\nu/2\gamma(r)$, and wide time range, since photons emitted
further from the source-observer line of sight arrive at later time and
are observed to have lower energy. There is no unique
relation, therefore, between the radius $r$ and the time or frequency 
at which radiation 
emitted at $r$ is observed. However, since most of the
emission from radius $r$ detected by a distant observer originates from a 
disk of radius $\sim r/\gamma(r)$ around the source-observer line of sight, 
and since photons emitted at a distance 
$r/\gamma(r)$ from the line of sight with frequency $\nu$ (in the shell frame)
are observed with frequency 
$\gamma(r)\nu$ and are delayed (compared to photons emitted on the line
of sight) by $r/2\gamma^2(r)c$, it is commonly 
assumed that at observed time $t=r/2\gamma^2(r)c$ the flux peaks at 
frequency $\nu'=\gamma(r)\nu(r)$,
where $\nu(r)$ is the frequency at which the emission peaks in the
shell rest frame.
Based on the same argument, it is assumed that
photons of frequency $\nu'=\gamma(r)\nu(r)$ arrive over a time scale
$\sim r/2\gamma^2(r)c$. 

It had recently been claimed (\cite{Sari}) that due to shock deceleration
the commonly used relation $t=r/2\gamma^2(r)c$ should be replaced by
$t=r/16\gamma^2(r)c$. Applying 
the suggested deceleration modification would make it difficult to 
reconcile observed afterglow time scales with blast wave model predictions.
This modification may have other implications. For example, shortly after
the suggestion by Goodman (1997), that local inter-stellar scintillation
may modulate the radio flux of GRB afterglows, radio variability on
time scale consistent with scintillation origin has been observed 
(\cite{Frail}). However, while the source size based on using the relation 
$t=r/16\gamma^2(r)c$ is $\sim4$ times larger than required
to allow for diffractive scintillation, implying that only refractive 
scintillation is likely (Goodman 1997), the observed large 
amplitude of radio variability
suggests that modulation, if due to scintillation,
should be produced by diffraction. Using the relation 
$t=r/2\gamma^2(r)c$, on the other hand, 
implies a source size which is compatible with that required for 
diffractive scintillation. Similarly, the larger size implied by the suggested
modification implies that it would be difficult to observe microlensing of 
the optical emission (\cite{lens}).

In \S2 we present a detailed analysis of the implications of the relativistic 
hydrodynamics of expanding blast waves to the observed afterglow. 
We find that the modification due to shock deceleration of the numerical
coefficient in the expression $t=r/2\gamma^2(r)c$, giving the
time at which the flux peaks at observed frequency $\nu'=\gamma(r)\nu(r)$, is
small. It should be emphasized, that the exact value
of the numerical coefficient depends not only on the hydrodynamics, but also
on the details of the electron energy distribution and on the magnetic field 
distribution behind the shock. 
Due to the lack of a theory describing these distributions, the exact value
of the numerical coefficient can not be determined. However, it can be shown
that the effect of deceleration is small. This is due to the fact that the
relation $t=r/16\gamma^2(r)c$ gives the delay of photons emitted from 
the fireball at radius $r$ from a point on the shock front on the line of 
sight, while most photons suffer longer delays, since they are 
emitted from a shell of finite thickness behind the shock, and from positions
off the line of sight. We show that at time 
$t=r/2\gamma^2(r)c$ the fireball appears on the sky as a narrow ring of radius
$r/\gamma(r)$ (This conclusion is independent of the details of electron and
magnetic field distribution). Our conclusions and their implications are
summarized in \S3.

\section{Emission from a relativistic expanding blast wave}

Let us consider a strong, spherical, ultra-relativistic shock wave expanding
into an ambient medium of uniform density. For a shock propagating with
a Lorentz factor $\Gamma\gg1$, the Lorentz factor, number density and energy 
density of shocked fluid at the shock discontinuity are $\gamma=\Gamma/
\sqrt{2}$, $n=4\gamma n_i$ and $e=4\gamma^2 n_i m_p c^2$  
respectively, where $n_i$ is the number density ahead of the shock
(e.g. \cite{BnM}). 
Observations indicate that the fraction of blast wave energy carried by 
electrons is not large, $\sim0.1$, and that the electron cooling time is long
compared to the dynamical time, i.e. to the blast wave expansion time
(\cite{AGWa},b). This implies that the energy lost to radiation is small,
and that the blast wave energy is approximately constant (so called ``adiabatic
blast wave''). Since the total energy in shocked particles is proportional
to $n_i r^3 \gamma (e/n)$, conservation of energy implies 
\begin{equation}
\gamma=\gamma_0\left({r\over r_0}\right)^{-3/2}, 
\label{gamma}
\end{equation}
where $r_0$ is some fiducial radius.

In fireball models of GRB afterglow it is assumed that the fractions of energy
carried by magnetic field and by electrons are time independent. Under this
assumption the magnetic field $B$ and the characteristic electron Lorentz 
factor $\gamma_e$
(in the shell frame) scale as $B\propto\gamma_e\propto\gamma$.
The observed radiation is produced by synchrotron emission of the shock
accelerated electrons. The characteristic synchrotron frequency in the
shell frame scales as $\nu\propto\gamma_e^2B\propto\gamma^3$. The number
of photons emitted as the shock propagates a distance $dr$ may be
obtained as follows. The number of radiating electrons scales as $r^3$,
and the number of photons each electron emits per unit time in the shell frame
is proportional to the magnetic field $B$. The time in the shell frame
over which the shock propagates a distance $dr$ is $dr/\gamma c$.
Thus, the number of photons emitted scales as $dN/dr\propto r^3B/\gamma$.
Using (\ref{gamma}) we therefore have
\begin{equation}
\nu=\nu_0 \left({r\over r_0}\right)^{-9/2},\quad 
{dN\over dr}=L^{-1}\left({r\over r_0}\right)^3,
\label{nu}
\end{equation}
where $L$ is a constant with dimensions of length.

The shock heated gas expands relativistically in its rest frame. Since the
time, measured in the rest frame, for the shock to expand to radius $r$ is 
$\sim r/\gamma c$, relativistic expansion in the shell frame implies that the
rest frame thickness of the shell of shock heated gas is $\sim r/\gamma$. 
Thus, in the observer frame most of the shocked gas, and most of the blast 
wave energy, are concentrated in a shell of thickness
$\Delta r=\zeta r/\gamma^2$,
%\begin{equation}
%\Delta r=\zeta r/\gamma^2,
%\label{zeta}
%\end{equation}
where $\zeta$ is some constant. We can obtain an estimate of $\zeta$ by
assuming that the density in the shocked shell is uniform. In this case
conservation of particle number implies $4\pi r^2\gamma\Delta r4\gamma n=
4\pi r^3n/3$, i.e. $\zeta=1/12$. In the self-similar solutions of
(\cite{BnM}), which give the spatial dependence of the hydrodynamic
variables, $90\%$ of the energy is concentrated in a shell of thickness
corresponding to $\zeta=1/7$.
It is implicitly assumed in fireball afterglow
models that the fractions of energy carried by electrons and magnetic field
vary behind the shock over a length scale comparable to the scale
for changes in the hydrodynamic variables, $\Delta r$. This is a reasonable
assumption, since the synchrotron cooling time of the electrons is longer than
the dynamical time, i.e. the characteristic expansion time. If the energy
fractions vary behind the shock on a scale much shorter than $\Delta r$, due
to some non hydrodynamic process which is not accounted for, a new
length scale would be introduced into the problem and the scalings (\ref{nu}),
on which the fireball afterglow model relies, will no longer hold. Since
the details of the spatial dependence of the electron and magnetic field 
energy fractions are not known, we will assume below that radiation
is produced within a homogeneous shell of width $\Delta r$ behind
the shock, and will derive results for different values of $\zeta$. 
For clarity we first assume that at a fixed time all emitted photons have the 
same energy (in the fireball frame). We then discuss modifications expected
due to a finite frequency range of emitted photons. 

In the shell frame photons are emitted isotropically. Consider, therefore, 
the appearance of an isotropic distribution of photons of frequency 
$\nu$ in a frame moving with Lorentz factor $\gamma$ with respect to the frame 
where the distribution is isotropic. We refer to the frame where the 
distribution is isotropic as the ``rest frame'' . 
Denoting with primes quantities measured in the moving frame, the rest
frame and moving frame frequencies are related by
\begin{equation}
\nu'=\gamma(1+\beta\cos\theta)\nu,
\label{nup}
\end{equation}
where $\beta=(1-1/\gamma^2)^{1/2}$ and $\theta$ is the (rest frame)
angle between the photon momentum and the direction of motion of the moving 
frame. The angle measured in the moving frame is
\begin{equation}
\tan\theta'={\sin\theta\over\gamma(\beta+\cos\theta)}.
\label{thetap}
\end{equation}
The fraction $df'$ of photons with frequencies in the range $\nu'$ to 
$\nu'+d\nu'$ is
\begin{equation}
{df'\over d\nu'}={1\over2}\sin\theta{d\theta\over d\nu'}=
{1\over2\gamma\beta\nu}\quad{\rm for}\quad\nu'_{\rm min}<\nu'<\nu'_{\rm max}.
\label{df}
\end{equation}
Here, $\nu'_{\rm min}=\gamma(1-\beta)\nu$, $\nu'_{\rm max}=\gamma(1+\beta)\nu$.

Let us now consider photons observed by a distant observer in a frequency
range $\nu'$ to $\nu'+d\nu'$. The number of photons produced
by the fireball shell at radius $r$, with observed frequency in the range 
$\nu'$ to $\nu'+d\nu'$, is given by 
(\ref{nu}) and (\ref{df})
\begin{equation}
{d^2N\over d\nu'dr}=L^{-1}{r^3\over2\gamma(r)\nu(r)r_0^3}= 
{1\over2L\gamma_0\nu_0}\left({r\over r_0}\right)^9,
\label{dN}
\end{equation}
where we have approximated $\beta=1$ in (\ref{df}), as we are interested in 
the limit $\gamma\gg1$. 
Any photon which arrives at a distant observer
must be emitted from the fireball in a direction parallel to the 
source-observer line of sight. A photon emitted at radius $r$, with frequency
$\nu(r)$ in the shell frame, is observed with frequency $\nu'$ provided,
therefore, it is emitted from the fireball on a line emerging from the 
explosion center and making an angle $\theta'$, given by (\ref{thetap})
and (\ref{nup}),
with the source-observer line of sight. Such a photon is delayed with respect
to photons emitted on the line of sight by
\begin{equation}
\Delta t_\theta=(1-\cos\theta'){r\over c}={1-\beta\over\beta}{r\over c}
\left[\gamma(1+\beta){\nu\over\nu'}-1\right]
\quad{\buildrel{\gamma\rightarrow\infty}\over\longrightarrow}\quad
{r\over2\gamma^2 c}\left(2\gamma{\nu\over\nu'}-1\right).
\label{t_theta}
\end{equation}
The total delay, i.e. the delay with respect to photons emitted from the
center of the explosion $r=0$, is given by the sum of $\Delta t_\theta$ and
$\Delta t_r=t_{\rm e}-r/c$, the difference between light travel time to radius 
$r$ and shock expansion time to radius $r$, 
$t_{\rm e}=\int_0^r(1-1/\Gamma^2)^{-1/2}dr/c$. 
For the scaling (\ref{gamma}) we have, for $\gamma\gg1$ and
using $\Gamma=2^{1/2}\gamma$, $\Delta t_r=r/16\gamma^2c$. Thus, 
the delay of photons emitted from the shock front at radius $r$, and observed 
with frequency $\nu'$, with respect to photons emitted from the center of the
explosion $r=0$, is 
\begin{equation}
t(\nu',r)=\Delta t_\theta(\nu',r)+r/16\gamma^2c.
\label{t_o}
\end{equation}
Since photons are emitted uniformly from a shell of finite 
thickness $\zeta r/\gamma^2c$ behind the shock, the arrival times of photons 
emitted at 
radius $r$ and observed with frequency $\nu'$ are uniformly distributed 
(for $\gamma\gg1$)
over the range $t=t(\nu',r)$ to $t=t(\nu',r)+\tau_\zeta$, where
$\tau_\zeta=\zeta r/\gamma^2c$.
%\begin{equation}
%\tau_\zeta=\zeta r/\gamma^2c.
%\label{t_zeta}
%\end{equation}

Using (\ref{dN}--\ref{t_o}) we have numerically calculated the 
intensity (energy flux per unit frequency) $f_\nu\equiv\nu'd^2N/d\nu'dt$ 
as a function of time. Results
are shown in Fig. 1 for different values of the shell thickness,
$\zeta=1/4,\ 1/16,\ {\rm and}\ 1/64$. Fig. 1 presents the time dependent
flux at $\nu'=\gamma_0\nu_0$. The intensity at other frequencies is obtained
using the scaling
\begin{equation}
f_\nu(\nu_2',t)=f_\nu\left[\nu_1',\left({\nu_2'\over\nu_1'}\right)^{-2/3}t
\right].
\label{scaling}
\end{equation}
For the shell width expected from fireball hydrodynamics, $\zeta\sim1/10$, the
intensity peaks at frequency $\nu'=\gamma(r)\nu(r)$ at a time
$t\simeq r/4\gamma^2(r)c$. The numeric coefficient in this relation, $1/4$,
is somewhat smaller than the commonly used value, $1/2$. However, as explained
above, since the spatial dependence of the fractions of energy carried by
electrons and magnetic field is not known, the effective shell width $\zeta$
can not be determined exactly (as it is not determined by the hydrodynamics
only), and therefore the value of the coefficient in the relation 
$t\simeq r/4\gamma^2(r)c$ can not be determined exactly. 
It is nevertheless clear 
that the intensity peaks at $\nu'=\gamma(r)\nu(r)$ at a delay significantly
larger than $t=r/16\gamma^2(r)c$.

The results presented in Fig. 1 were obtained under the assumption that 
at a fixed shell radius all emitted photons have the same frequency $\nu(r)$ 
(in the fireball frame). In fireball afterglow models the emission peaks
at $\nu(r)$, and extends to high frequency $\nu\gg\nu(r)$ due to power-law 
energy distribution of electrons,
and to low frequency $\nu\ll\nu(r)$ due to the low energy tail of
synchrotron emission, $f(\nu)/f[\nu(r)]\sim[\nu/\nu(r)]^{1/3}$. These tails
dominate the intensity at observed frequency $\nu'=\gamma(r)\nu(r)$ at delays
much smaller and much larger than the time at which the intensity peaks 
at $\nu'$, $t\ll r/\gamma^2c$ and $t\gg r/\gamma^2c$. However, they are
not important near the peak, $t\sim r/\gamma^2c$. The details of the
frequency dependence of the emission at $\nu\sim\nu(r)$ will affect the 
behavior at $t\sim r/\gamma^2c$. Let us assume that photons are 
emitted at the shell frame at two 
different frequencies, $\nu(r)$ as given in (\ref{nu}) and 
$\tilde\nu(r)=x\nu(r)$. It is straight forward to show that the
intensity $\tilde f_\nu(\nu',t)$ due to emission at $\tilde\nu(r)$ is 
related to that of the intensity $f_\nu(\nu',t)$ due to emission at 
$\nu(r)$ by $\tilde f_\nu(\nu',t)\propto f_\nu(\nu',x^{-2/3}t)$. 
Thus, if most of the energy is emitted in the shell frame over a frequency
range $\delta\nu(r)/\nu(r)\sim1$, the frequency spread would introduce a spread
in arrival time of photons of frequency $\nu'$, $\delta t/t_p\sim1$ where
$t_p$ is the time at which the intensity peaks under the assumption that
all photons are emitted at the same frequency $\nu(r)$. From Fig. 1, the
spread in arrival time due to the relativistic expansion is $\delta t/t_p
\sim2$. Thus, the effect of emission over a finite frequency range would
not be significant, provided most of the energy is emitted over a frequency
range $\delta\nu(r)/\nu(r)\lesssim1$.

The flux observed at a fixed time originates from points at a range of
distances transverse to the line of sight. 
Using the equations derived above we have numerically calculated the 
fraction of flux contributed from rings of radii $h$--$h+dh$
around the line of sight, as function of $h$. 
The results are shown in Fig. 2 for several $\zeta$
values. At time $t=r/2\gamma^2(r)c$ the flux originates from a ring of outer
radius $h\simeq r/\gamma(r)$ and width $\Delta h/h\sim0.1$. 
Note that this result is independent of the spectral distribution of the
emission in the shell frame. The appearance of the fireball as a narrow ring
can be qualitatively understood from Fig. 3. 
Radiation produced
at $r_0$ and observed at $t_0=r_0/2\gamma^2(r_0)c$ originates from an
angle $\sim1/\gamma_0$, corresponding to $h=r_0/\gamma_0$. Contribution to
the emission at
smaller $h$ is due to emission from $r<r_0$ and $r>r_0$. However,
the contribution from smaller radii is suppressed due to the fact that
at smaller radii the radiation seen at a given time originates from larger 
angles, and the relativistic beaming suppresses the emission from angles
$>1/\gamma$. The contribution from larger radii is small since the emissivity
decreases as the fireball decelerates.

\section{Conclusions}

We have presented a detailed analysis of the implications of the relativistic
hydrodynamics of expanding blast waves to GRB afterglow observations. We have
shown that the afterglow intensity peaks at observed frequency  
$\nu'=\gamma(r)\nu(r)$, where $\nu(r)$ is the frequency at which the emission 
peaks in the fireball frame at radius $r$, at time $t\simeq r/4\gamma^2(r)c$.
The exact value of the numerical coefficient depends on the effective thickness
of the radiating shell, $\zeta r/\gamma^2$. The value $1/4$ is
obtained for $\zeta\sim1/10$, the value implied by fireball hydrodynamics
(see Fig. 1). The numeric coefficient, $1/4$,
is somewhat smaller than the commonly used value, $1/2$. However, it should
be kept in mind that since the spatial dependence of the fractions of energy 
carried by electrons and magnetic field is not known, the effective shell 
width $\zeta$ can not be determined exactly, and the numerical coefficient 
may be somewhat larger or smaller then $1/4$. The amplitude of the peak 
intensity agrees with the estimate of (\cite{AGWa},b). For
the intensity normalization chosen in Fig. 1, the peak intensity estimated 
following (\cite{AGWa},b) is $2/3$, close to the numerical results obtained 
here. This is due to the fact that most photons of frequency 
$\nu'=\gamma(r)\nu(r)$ arrive
over a time scale $t\simeq r/2\gamma^2(r)c$, as assumed in (\cite{AGWa},b).

At time $t$ the fireball
appears on the sky as a narrow ring of radius $h=r/\gamma(r)$ and width 
$\Delta h/h\sim0.1$, where $r$ and $t$ are related by $t=r/2\gamma^2(r)c$
(see Fig. 3). The apparent size $h=r/\gamma(r)$ is smaller
by a factor of $(16/2)^{5/8}\sim4$ compared to that obtained by using 
the relation $t=r/16\gamma^2(r)c$ (this follows
from [\ref{gamma}]). The smaller 
size implies that diffractive scintillation is likely to modulate the
afterglow radio flux (\cite{Goodman}), and that significant modification 
of the optical light curve due to microlensing is possible
on day time scale (\cite{lens}). The narrowness of the emission ring 
would affect the predictions for both microlensing and scintillation (This
has been taken into account in the lensing calculations of [\cite{lens}].
More detailed calculations are required for modulation by scintillation).

The results presented here are valid for highly 
relativistic fireballs, $\gamma\gg1$. For Lorentz factors $\gamma-1\sim1$,
the emission ring would be wider. Finally, we note that although we have
implicitly assumed spherical symmetry throughout the paper, our results
are valid for a fireball which is a cone of finite opening angle $\theta$
as long as $\gamma>1/\theta$. This is due to the fact that most of the 
observed emission originates from a cone of opening angle $1/\gamma$.

\paragraph*{Acknowledgments.} 

I thank J. N. Bahcall for helpful discussions. This research
was partially supported by a W. M. Keck Foundation grant 
and NSF grant PHY95-13835.

\newpage

\begin{figure}
\plotone{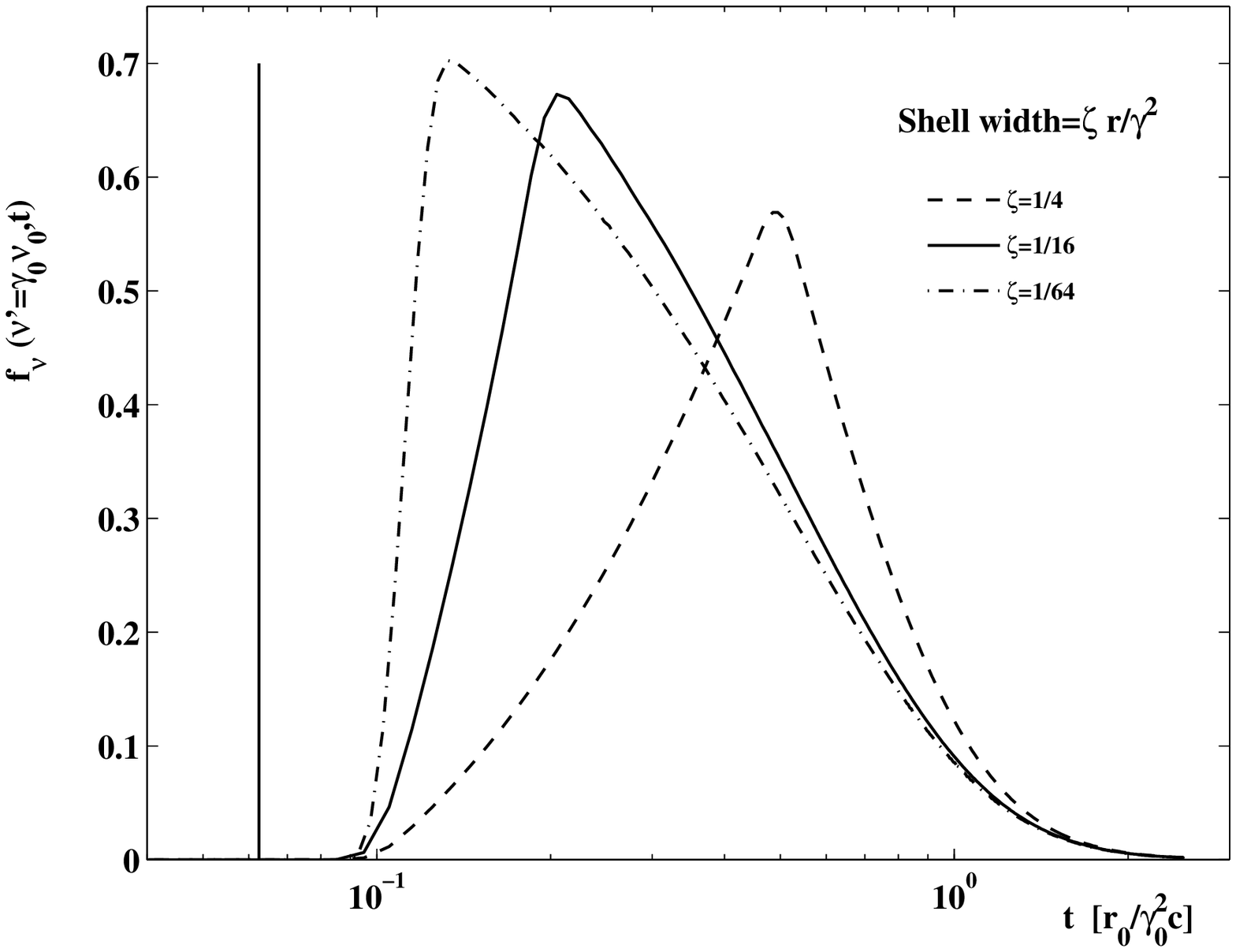}
\caption{
The observed intensity (energy flux per unit frequency) at observed
frequency $\nu'=\gamma(r_0)\nu(r_0)$ as a function of time for various 
fireball shell widths. The width expected from hydrodynamic considerations 
corresponds to $\zeta\sim1/10$. $\nu(r_0)$
is the frequency with which photons are emitted in the shell frame at
$r=r_0$. The normalization of the intensity is determined by measuring
time in units of $r_0/\gamma^2(r_0)c$, and by setting the dimensional
coefficient in (\ref{nu}) to $L=1$.
The vertical solid line denotes the time $t=r_0/16\gamma^2(r_0)c$.
}
\label{fig1}
\end{figure}

\begin{figure}
\plotone{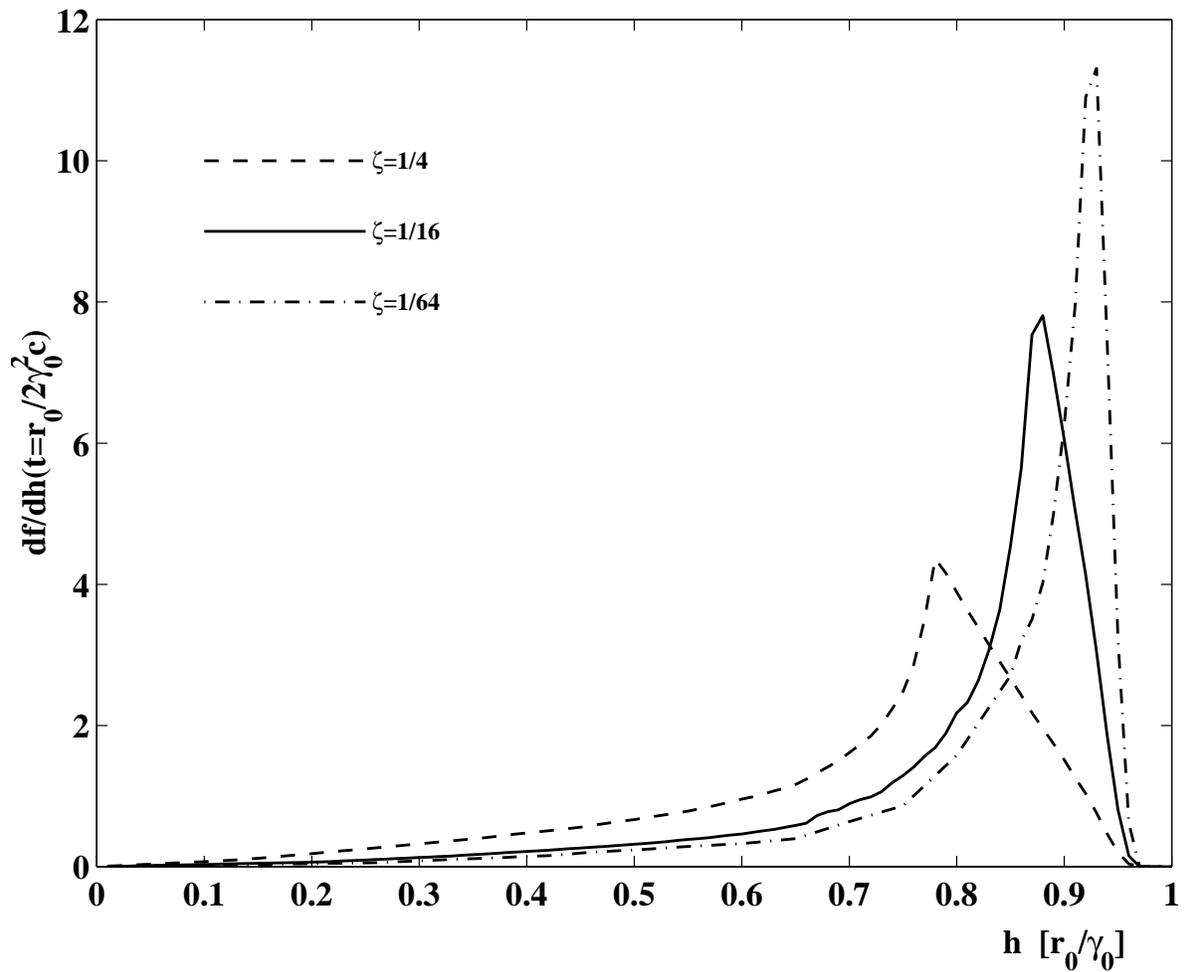}
\caption{
The fraction of flux observed at time $t=r_0/2\gamma^2(r_0)c$ 
which originates from different distances $h$ transverse to the line of sight,
for various fireball shell widths $\zeta r/\gamma^2(r)$.
}
\label{fig2}
\end{figure}

\begin{figure}
\plotone{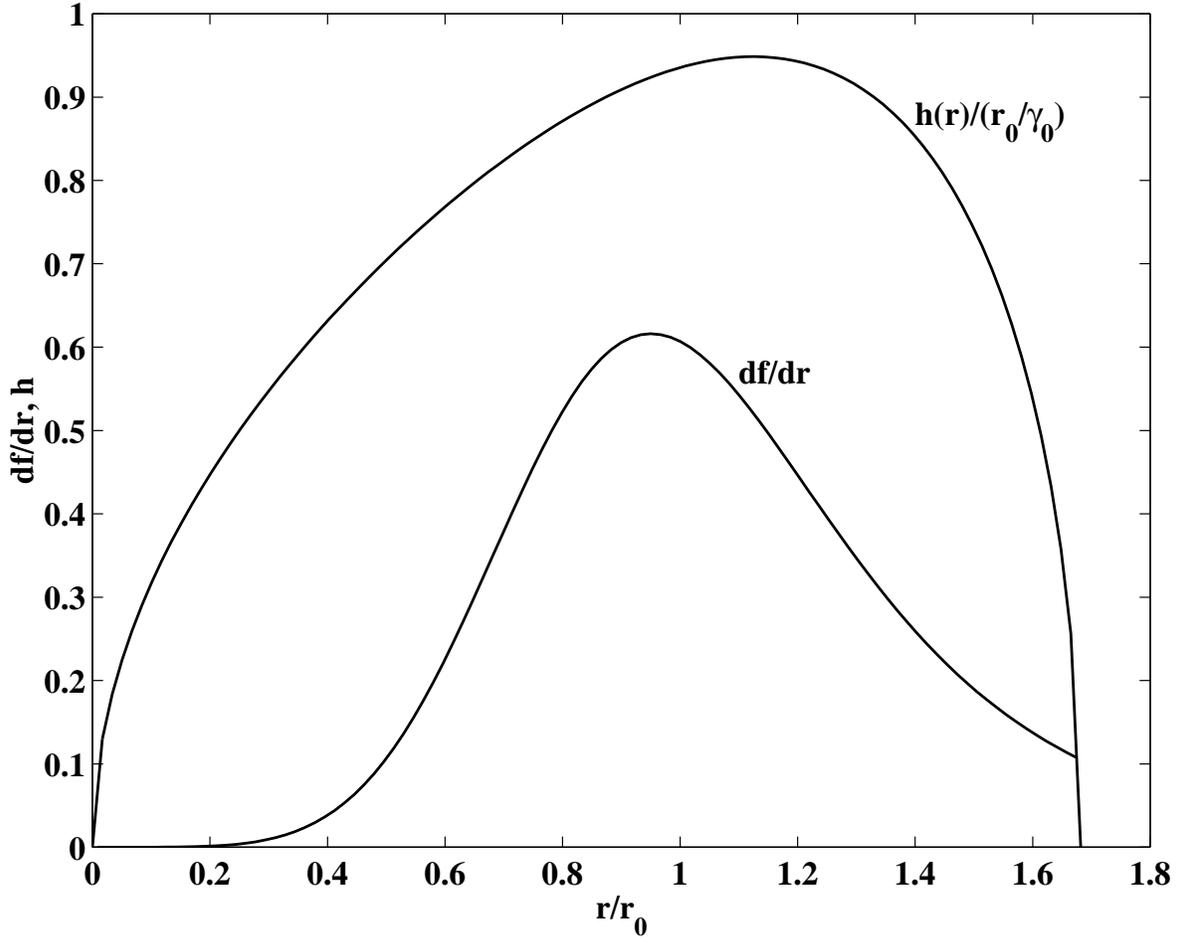}
\caption{
The fraction of flux obserevd at time $t=r_0/2\gamma^2(r_0)c$ 
and produced over a radii range $dr$, $df/dr$, plotted as
a function of $r$ with $h(r)$, the distance transverse to the line of
sight from which radiation emitted at $r$ arrives at the observer at time $t$,
for a shell of thickness $\zeta\rightarrow0$.
}
\label{fig3}
\end{figure}

\end{document}